\documentclass[twocolumn]{aastex63}

\usepackage{rotating}
\usepackage{graphicx}
\usepackage{appendix}
\usepackage{float}
\usepackage{hyperref}
\usepackage{amssymb,amsmath}

\newcommand{\p}[1]{{\color{magenta}{#1}}}

\renewcommand{\d}[1]{{\color{red}{#1}}}

\newcommand{\angstrom}{\mbox{\normalfont\AA}}
\newcommand{\BE}{\begin{equation}}
\newcommand{\EE}{\end{equation}}
\newcommand{\BA}{\begin{align}}
\newcommand{\EA}{\end{align}}

\received{XXXX, 2020}
\revised{XXXX, 2020}
\accepted{November 15, 2020}
\submitjournal{AJ}

\shortauthors{et al.}
 \watermark{draft}
\graphicspath{{./}{figures/}}

\begin{document}

\title{Alfv\'enic Perturbations in a Sunspot Chromosphere Linked to Fractionated Plasma in the Corona}

\author[0000-0002-0665-2355]{Deborah Baker}
\affiliation{University College London, Mullard Space Science Laboratory, Holmbury St. Mary, Dorking, Surrey, RH5 6NT, UK}

\author[0000-0002-5365-7546]{Marco Stangalini}
\affiliation{ASI Italian Space Agency, Via del Politecnico, s.n.c
00133 – Roma, Italia}

\author[0000-0001-7809-0067]{Gherardo Valori}
\affiliation{University College London, Mullard Space Science Laboratory, Holmbury St. Mary, Dorking, Surrey, RH5 6NT, UK}

\author[0000-0002-2189-9313]{David H. Brooks}
\affiliation{College of Science, George Mason University, 4400 University Drive, Fairfax, VA 22030, USA}

\author[0000-0003-0774-9084]{Andy S.H. To}
\affiliation{University College London, Mullard Space Science Laboratory, Holmbury St. Mary, Dorking, Surrey, RH5 6NT, UK}

\author[0000-0002-2943-5978]{Lidia van Driel-Gesztelyi}
\affiliation{University College London, Mullard Space Science Laboratory, Holmbury St. Mary, Dorking, Surrey, RH5 6NT, UK}
\affiliation{LESIA, Observatoire de Paris, Universit\'e PSL, CNRS, Sorbonne Universit\'e, Univ. Paris Diderot, Sorbonne Paris Cit\'e, 5 place Jules Janssen, 92195 Meudon, France}
\affiliation{Konkoly Observatory, Research Centre for Astronomy and Earth Sciences, Hungarian Academy of Sciences, Konkoly Thege \'ut 15-17., H-1121, Budapest, Hungary}
\author[0000-0001-8215-6532]{Pascal D\'emoulin}
\affiliation{LESIA, Observatoire de Paris, Universit\'e PSL, CNRS, Sorbonne Universit\'e, Univ. Paris Diderot, Sorbonne Paris Cit\'e, 5 place Jules Janssen, 92195 Meudon, France}

\author[0000-0002-1365-1908]{David Stansby}
\affiliation{University College London, Mullard Space Science Laboratory, Holmbury St. Mary, Dorking, Surrey, RH5 6NT, UK}

\author[0000-0002-9155-8039]{David B. Jess}
\affiliation{Astrophysics Research Centre, School of Mathematics and Physics, Queen's University Belfast, Belfast, BT7 1NN, UK}
\affiliation{Department of Physics and Astronomy, California State University Northridge, Northridge, CA 91330, U.S.A.}

\author[0000-0002-7711-5397]{Shahin Jafarzadeh}
\affiliation{Rosseland Centre for Solar Physics, University of Oslo, P.O. Box 1029 Blindern, NO-0315 Oslo, Norway}
\affiliation{Institute of Theoretical Astrophysics, University of Oslo, P.O. Box 1029 Blindern, NO-0315 Oslo, Norway}

\begin{abstract}
 In this study, we investigate the spatial distribution of highly varying plasma composition around one of the largest sunspots of solar cycle 24. 
 Observations of the photosphere, chromosphere, and corona are brought together with magnetic field modelling of the sunspot in order to probe the conditions which regulate the degree of plasma fractionation within loop populations of differing connectivities.
 We find that in the coronal magnetic field above the sunspot umbra, the plasma has photospheric composition. 
Coronal loops rooted in the penumbra contain  fractionated plasma, with the highest levels observed in the loops that connect within the active region.  
Tracing field lines from regions of fractionated plasma in the corona to locations of Alfv\'enic fluctuations detected in the chromosphere shows that they are magnetically linked. 
These results indicate a connection between sunspot chromospheric activity and observable changes in coronal plasma composition. 
\end{abstract}


\section{Introduction} \label{intro}
Early observations of elemental abundance variations on the Sun showed systematic differences between the composition of the corona and that of the photosphere \citep[e.g.][]{widing89,widing95,sheeley95,sheeley96}.
In the closed-loop solar corona, in the slow solar wind, and in solar energetic particles (SEPs), elements with low first ionization potential (FIP $<$10 eV) are more abundant by a factor of 2--4 compared to the photosphere  \citep[e.g.][]{meyer85a,meyer85b,gloeckler89,feldman03,brooks15}, whereas, high FIP elements (FIP $>$ 10 eV) retain their photospheric elemental distribution.
Plasma composition in the open magnetic field of coronal holes remains relatively unfractionated when it is observed in the corona \citep[e.g.][]{feldman93,feldman98,doschek98,brooks11a}.
Abundance variations are typically characterized by FIP bias which is the ratio of an element's abundance in the solar atmosphere to its abundance in the photosphere.
FIP bias of $\sim$1 indicates unfractionated photospheric plasma composition and $>$1.5 is fractionated plasma of coronal composition.

\cite{feldman90} provided one of the few early studies of plasma composition around a sunspot based on spatially unresolved, slit observations obtained from a rocket flight of the High Resolution Telescope and Spectrograph (HRTS).
The authors determined that in the atmosphere above a sunspot, the elemental abundances had a photospheric distribution compared to the plasma of the nearby plage region, which was highly enriched in low FIP elements.
Similarly, \cite{sheeley95} noted in a \emph{Skylab} slitless spectrogram that plasma above a sunspot umbra was enriched in high-FIP \ion{Ne}{6} whereas in the adjacent penumbra, the plasma was enriched in low-FIP \ion{Mg}{6}.
\ion{Ne}{6} rich plasma occurred only in areas of flux emergence in two nearby active regions.

Subsequent studies of high FIP bias plasma in active regions refer to features such as \ion{Mg}{9} sprays \citep{sheeley96}, spikes at the edges of active regions \citep{young97a},
fan loops \citep{warren16}, and upflow/outflow regions \citep[e.g.][]{brooks11a}.
Typically, the magnetic field associated with these features is the decaying or dispersed unipolar areas of strong magnetic field at the periphery of active regions.
Strong plasma fractionation is observed at the footpoints of loops rooted in the unipolar regions where FIP bias levels are 3--4 \citep{brooks11a,baker13,brooks15}.
High FIP bias of $\sim$3 is also observed in the cores of quiescent active regions \citep{delzanna14}.

According to the plasma fractionation model of \cite{laming15}, a compelling explanation for the separation of ions from neutrals is the ponderomotive force arising from the reflection or refraction of Alfv\'en waves in the chromosphere.
The Alfv\'en waves act only on the ions while leaving neutral elements unaffected.
Though the fractionation is influenced by the origin and flux of the Alfv\'en waves as well as the wave-wave interactions in the chromosphere, in general, the time averaged ponderomotive force is directed upward, giving rise to the enrichment of easy-to-ionize low FIP elements in the corona \citep{laming15,laming17}.
The direction and ultimately the resonance of the Alfv\'en waves are all-important to the degree of fractionation observed in the corona  \citep{laming19}. 
These features are set by where the Alfv\'en waves are generated.
In open field regions, typical waves with 3 and 5 min periods \citep[e.g.][and references therein]{khomenko15} generated from below the photosphere propagate upward at the base of the field and either continue along the open field or are reflected back down; there is little resonance, therefore, little fractionation.
Upward propagating waves with such long periods do not resonate with the closed loop corona so like with open field regions, the waves are reflected back down at the loop footpoints resulting in little or no fractionation.
Conversely, Alfv\'en waves generated in the corona due to magnetic reconnection are directed downward to loop footpoints at the top of the chromosphere and then are reflected back upward at the steep density gradient located there.
\cite{laming17} proposed that resonant waves are excited within the coronal loop itself as a result of nanoflare reconnection in the corona thereby creating enhanced fractionation at magnetically connected loop footpoints \citep{baker13,dahlburg16,laming17,laming19}.
It is not observationally clear if these oscillations generated by reconnection in the corona are linked to enhanced fractionation, however.

In this regard, the quest for magnetic fluctuations associated with magneto-hydrodynamic waves (MHD) in solar magnetic structures assumes a particular importance.  Observationally, MHD waves in solar magnetic structures are generally detected as intensity and velocity oscillations \citep{bogdan00, centeno06,chorley10,morton11,stangalini12,grant15,jafarzadeh17, jess17}, although simultaneous magnetic fluctuations are also expected from theory for different types of MHD modes \citep{Edwin1983,roberts83}.
The required magnetic oscillations for ponderomotive fractionation to take place can therefore be associated with a number of different wave modes \citep{roberts83,khomenko03,goossens09,morton15} e.g. locally excited waves (Alfv\'en, magneto-acoustic fast mode in high-$\beta$ regimes), or global eigenmodes of the magnetic structure (e.g. sausage mode, torsional Alfvén mode,...). 
We will refer to magnetic fluctuations associated with any wave mode as Alfv\'enic waves, to distinguish from a purely Alfvén mode.
The detection of magnetic oscillations associated with MHD modes is a difficult task as opacity effects or instrumental crosstalk with other physical quantities can easily mimic the effect of these oscillations \citep[e.g.][and references therein]{khomenko15,joshi18}. 
One way to disentangle intrinsic magnetic oscillations in the solar atmosphere is through the investigation of the phase lag between the polarization signals associated with magnetic field disturbances and other physical quantities such as intensity and Doppler velocity \citep{stangalini18, stangalini20}.
Oscillating physical quantities associated with MHD waves may have different phase relations depending on the MHD mode and the propagation state of the wave \citep{fujimura09, moreels13a, moreels13b, hinode19}, thus the analysis of the phase relations between them can be exploited for their identification \citep{stangalini18}, and the identification of the specific mode producing them.

In this paper, we show a detailed, spatially resolved coronal  composition map of a strong and coherent leading sunspot in AR 12546.
We find that the elemental abundance variation in the corona above the sunspot is highly structured with extremes in the level of fractionation among the distinct loop populations.
The distribution of the highly fractionated plasma appears correlated with the spatial locations at which intrinsic magnetic oscillations are identified in nearly simultaneous high spatial resolution spectropolarimetric observations of the solar chromosphere \citep{stangalini20}. 
Magnetic field modeling is used to investigate the connectivities of the loop populations within the sunspot to seek an understanding of the distribution of plasma composition observed there.
We interpret our findings in the wider context of coronal heating and the ponderomotive force model of elemental fractionation \citep{laming15}.
\begin{figure*}[ht]
    \centering
    \includegraphics[width=0.95\textwidth]{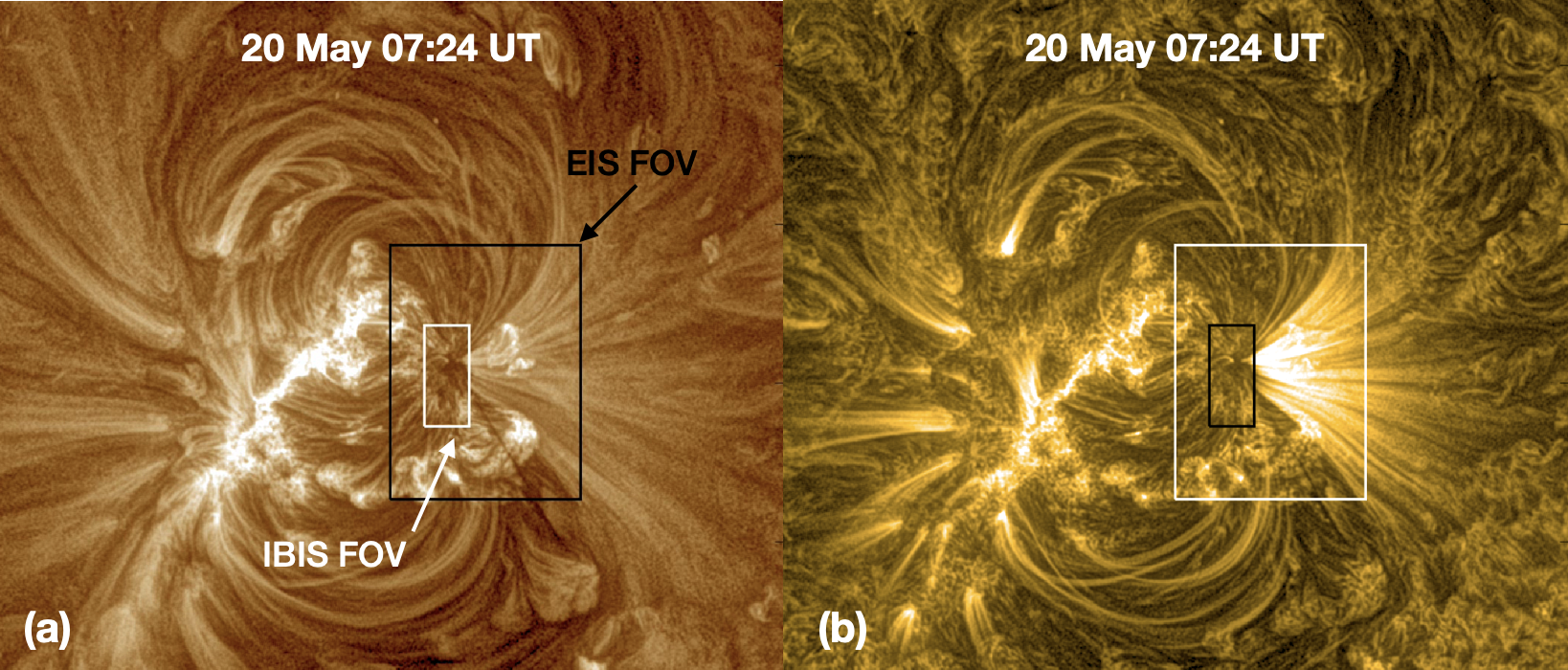}
    \includegraphics[width=0.95\textwidth]{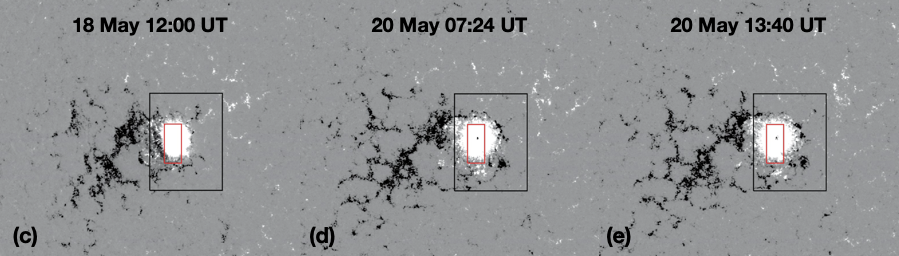}  
    \caption{\emph{SDO}/AIA 193 $\angstrom$ (a) and 171 $\angstrom$ (b) images at the time of the EIS observation at 07:24 UT on 2016 May 20.  \emph{SDO}/HMI magnetograms on May 18 at 12:00 UT (c), May 20 at 07:24 UT (d), and May 20 at 13:40 UT (e).  All images are overlaid with boxes showing the larger \emph{Hinode}/EIS (black) and smaller IBIS (red) FOVs. At 07:24 UT,
    the \emph{SDO}/AIA and HMI FOV is X = [-300$\arcsec$, 200$\arcsec$] and Y = [-280$\arcsec$, 140$\arcsec$].  Images of the figure are from the included movie labeled as fig1anim.mp4.  It covers the time period from 12:00 UT on May 18 to 14:55 UT on May 20 and lasts 30 s.
    \label{fig_context}}
\end{figure*}

\section{Observations and Methods} \label{obs}

\subsection{Overview of AR 12546} \label{overview}
The observations of AR 12546 were obtained on 2016 May 20 using the \emph{Hinode} EUV Imaging Spectrometer \citep[EIS;][]{culhane07}, the Interferometric BIdimensional Spectrometer at the National Solar Observatory Dunn Solar Telescope \citep[IBIS;][]{cavallini06,reardon08}, the Solar Dynamics Observatory (\emph{SDO})/Atmospheric Imaging Assembly \citep[AIA;][]{lemen12} and \emph{SDO}/Helioseismic and Magnetic Imager \citep[HMI;][]{scherrer12}.
Figure \ref{fig_context} shows high resolution \citep{morgan14} \emph{SDO}/AIA 193 $\angstrom$ (a) and 171 $\angstrom$ (b) context images centered on the active region overlaid with contours indicating the positions of the EIS and IBIS fields of view (FOVs).
At the time of the EIS raster, the sunspot is located at the solar central meridian (CM) approximately 100$\arcsec$ south of the equator.
The bottom panel of the figure contains a series of \emph{SDO}/HMI magnetograms (c--e) also with the EIS and IBIS FOVs.
Figure \ref{fig_context} includes an animation of these observations for 2016 May 18--20.

AR 12546, one of the largest ARs of the last 20 years, was a relatively simple, bipolar region composed of a strong, coherent leading positive polarity sunspot and a dispersed following field of negative polarity.
Asymmetric flux concentrations are typical of bipolar regions, however, both the extent of the dispersion of the following polarity and the coherency of the leading spot are extreme in this case.
At the time of its CM crossing on May 20, half of the total unsigned magnetic flux of the active region was 4.1$\times$10$^{22}$ Mx and the magnetic field strength was exceptionally high, exceeding 4,000 G in the center of the sunspot umbra in the photosphere \citep{stangalini18}.
There was no significant evolution of the large-scale field during the two days prior to the EIS and IBIS observations; the sunspot was globally stable. 
Small-scale evolution was limited to moving magnetic features streaming radially from the positive-polarity sunspot and ongoing fragmentation and dispersal of the negative field of the following polarity in the decaying active region \citep[see also][]{murabito19}. 

During the period of May 18--20, there were no flares $>$ B-class or coronal mass ejections (CMEs) attributable to AR 12546.
The stability of the loops rooted in the sunspot umbra and penumbra and the lack of activity reflect the absence of consequential evolution in the magnetic field (see the included animation of Figure \ref{fig_context}).

\begin{figure}[ht]
 \centering
    \includegraphics[trim=50 0 0 0, width=0.5\textwidth]{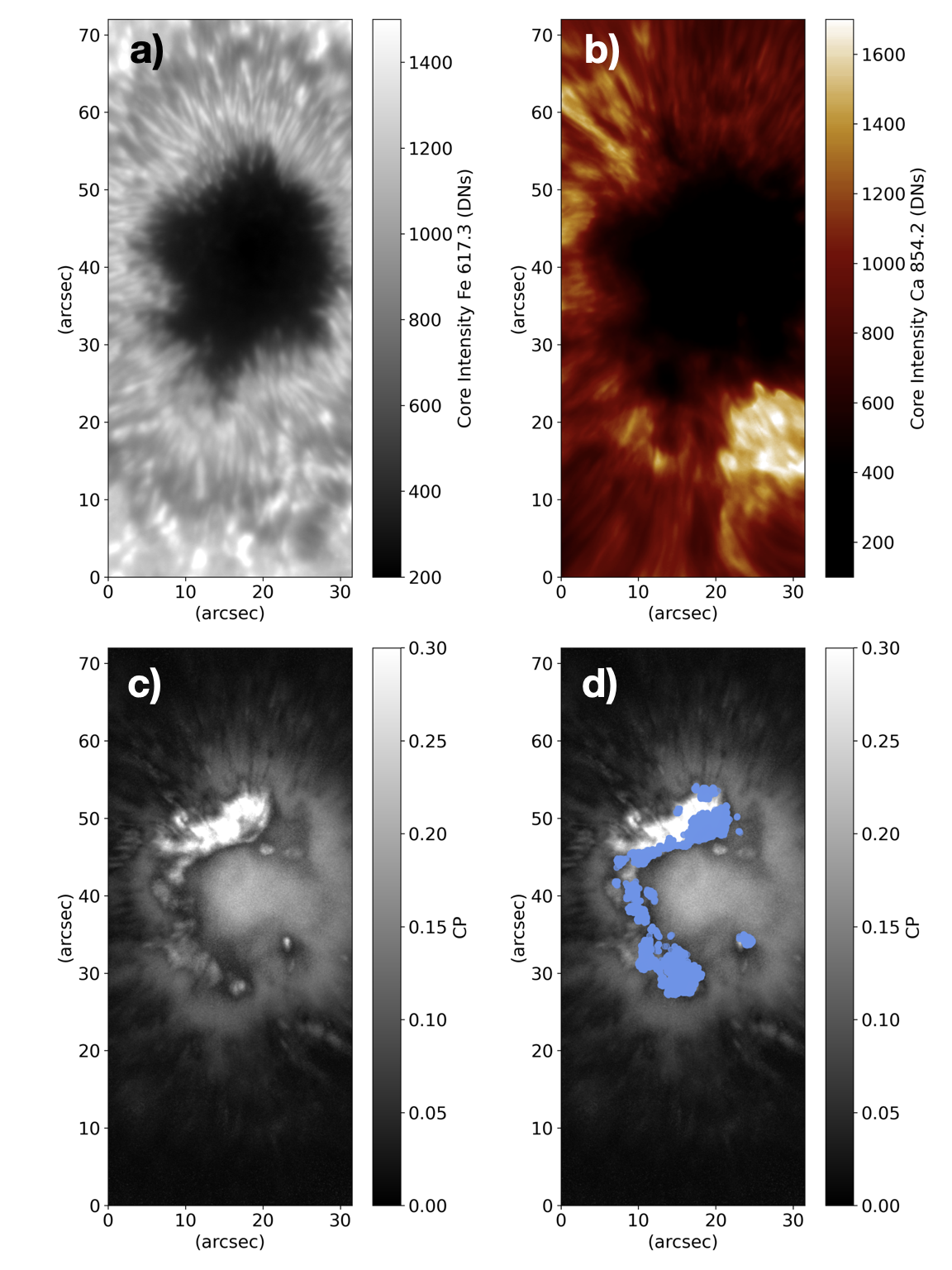}
    \caption{Top:  Photospheric IBIS \ion{Fe}{1} 6173 $\angstrom$ (a) and chromospheric  \ion{Ca}{2} 8542 $\angstrom$ core intensity (b) images at 13:39 UT on 20 May.
    Bottom panel:   Stokes \emph{V$_{max}$}/\emph{I$_{cont}$} $CP$ map without/with (c/d) blue dots overplotted. 
    The blue dots indicate the locations where the magnetic waves are detected (cf. Figures \ref{fig_fip}, \ref{fig_dots_match} and see the discussion  in Section \ref{sec_alfven}). The $CP$ map is saturated at 0.3 (white). 
    }
    \label{fig_ibis}
\end{figure}

\subsection{IBIS Observations and Methods}\label{sec_ibis}
IBIS full Stokes spectropolarimetric scans were used to identify possible signatures of magnetic field oscillations in circular polarization ($CP$) measurements in the umbra. 
A full account of the observations and data reduction techniques is provided in \cite{stangalini18,murabito19,houston20}, and \cite{stangalini20}.
Here we precis the aspects which are relevant to this analysis.
The data set consists of a time series of \ion{Ca}{2} 8542 $\angstrom$ scans beginning at 13:39 UT on May 20 and continuing for 184 minutes at a cadence of 48 sec. 
The \ion{Ca}{2} is a chromospheric magnetically sensitive line and therefore suitable for detecting magnetic field oscillations at chromospheric heights.
Figure \ref{fig_ibis} shows an IBIS intensity image in the  photospheric \ion{Fe}{1} 6173 $\angstrom$ line for context (a) and an intensity image in the \ion{Ca}{2} 8542 $\angstrom$ line (b).
The IBIS FOV of 28$\arcsec$$\times$70$\arcsec$ encompasses the umbra in the X-direction and a significant portion of the penumbra in the Y-direction.
The $CP$ measurements were obtained from the amplitude of the Stokes-\emph{V} profile.   $CP$ is defined as follows:
\begin{equation}\label{eq_cp}
CP = \frac{\lvert V_{max}\rvert}{I_{cont}} \cdot sign{\left(V_{max}\right)},
\end{equation}
where \emph{V$_{max}$} is the maximum amplitude of the Stokes \emph{V} spectral profile and I$_{cont}$ is the local continuum intensity \citep{stangalini18}.

To identify possible intrinsic magnetic field oscillations in the same IBIS data set, \cite{stangalini20} performed a specific phase lag analysis, between the $CP$ and the core intensity of the \ion{Ca}{2}. 
The Figure \ref{fig_ibis} (c) shows the Stokes \emph{V$_{max}$}/\emph{I$_{cont}$} $CP$ map saturated at 0.3.
The authors selected the $CP$ instead of Doppler velocity of the \ion{Ca}{2} line for the phase lag analysis due to the presence of shocks that can turn the \ion{Ca}{2} line from absorption to emission, rendering the line Doppler velocity undefined.
The $CP$ is directly related to line-of-sight magnetic field but its variation can be caused by either intrinsic magnetic oscillations or opacity effects.
In the case of opacity effects, the intensity is expected to be in/out of phase due to the fluctuation of the line formation height. 
This does not mean that real magnetic waves with the same phase relations (i.e. $\pm$$\pi$) do not exist, but the phase lag analysis provides a high level of confidence we are not observing a mix of real and unreal oscillations. 
A strong coherence is needed to ensure that unreliable phase measurements are excluded from the analysis.
In this work we make use of the results of \cite{stangalini20} and investigate the spatial distribution of the $CP$ oscillations with the aim of assessing their role in the distribution of the observed FIP bias in and around the sunspot.

\begin{figure*}[ht]
\centering
\includegraphics[width=0.18\textwidth]{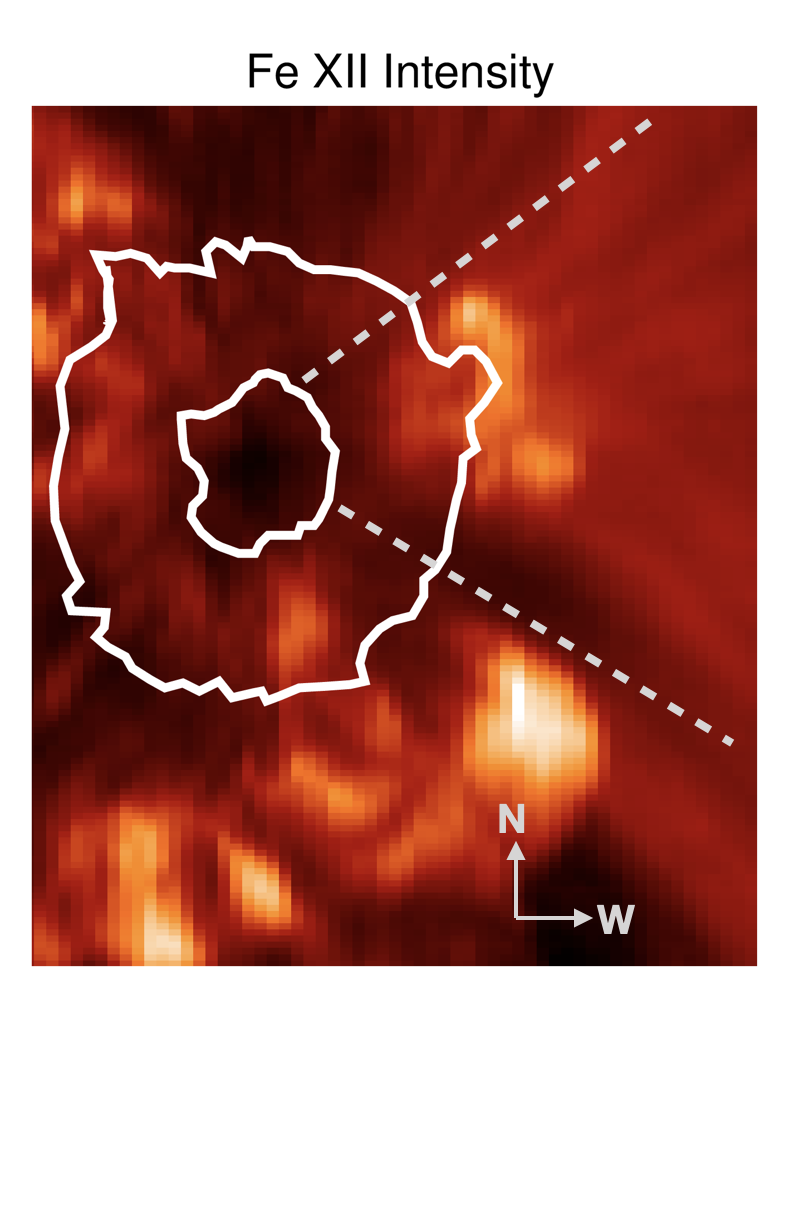}
\includegraphics[width=0.18\textwidth]{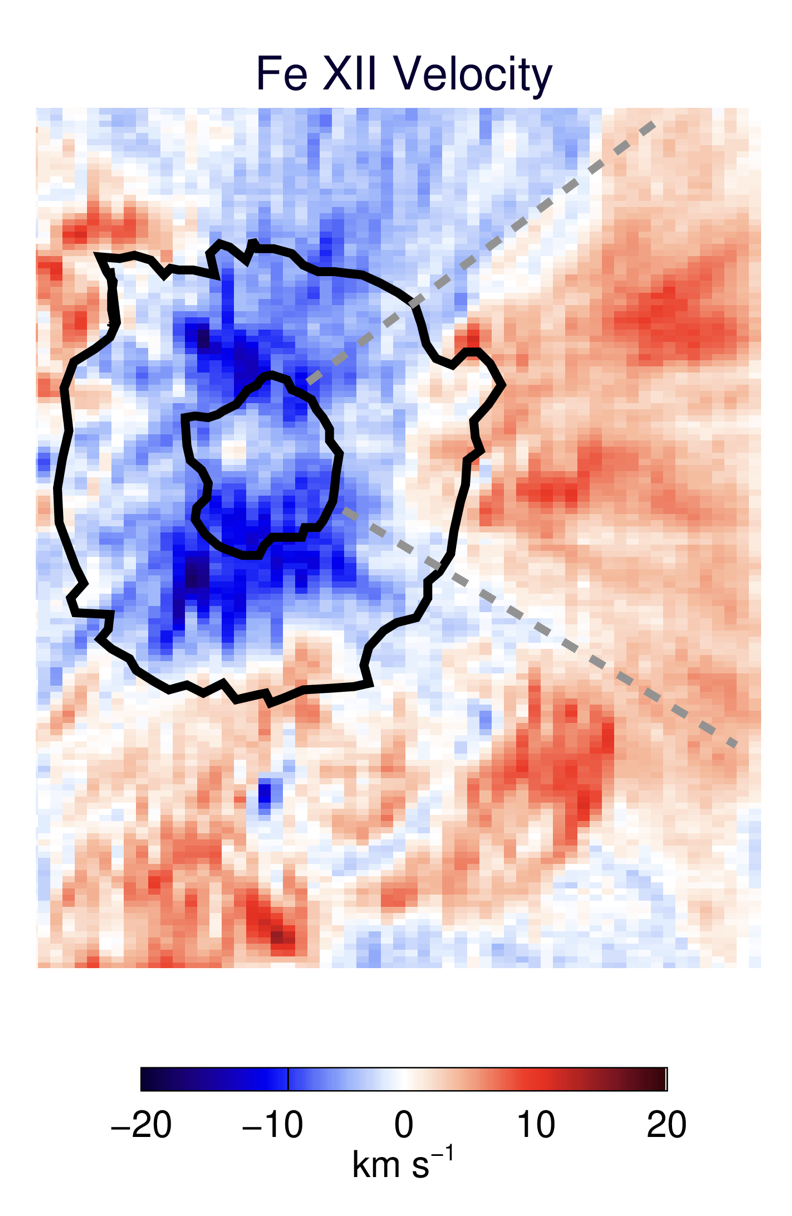}
\includegraphics[width=0.18\textwidth]{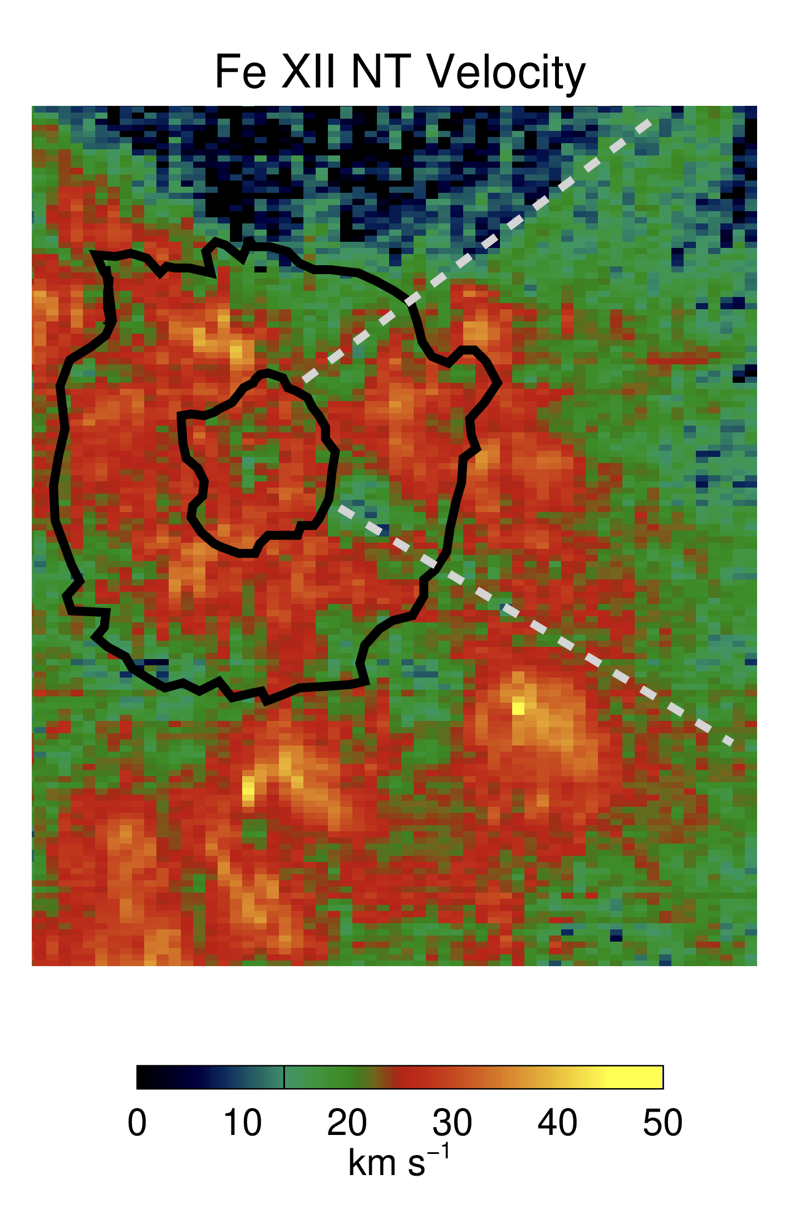}
\includegraphics[width=0.18\textwidth]{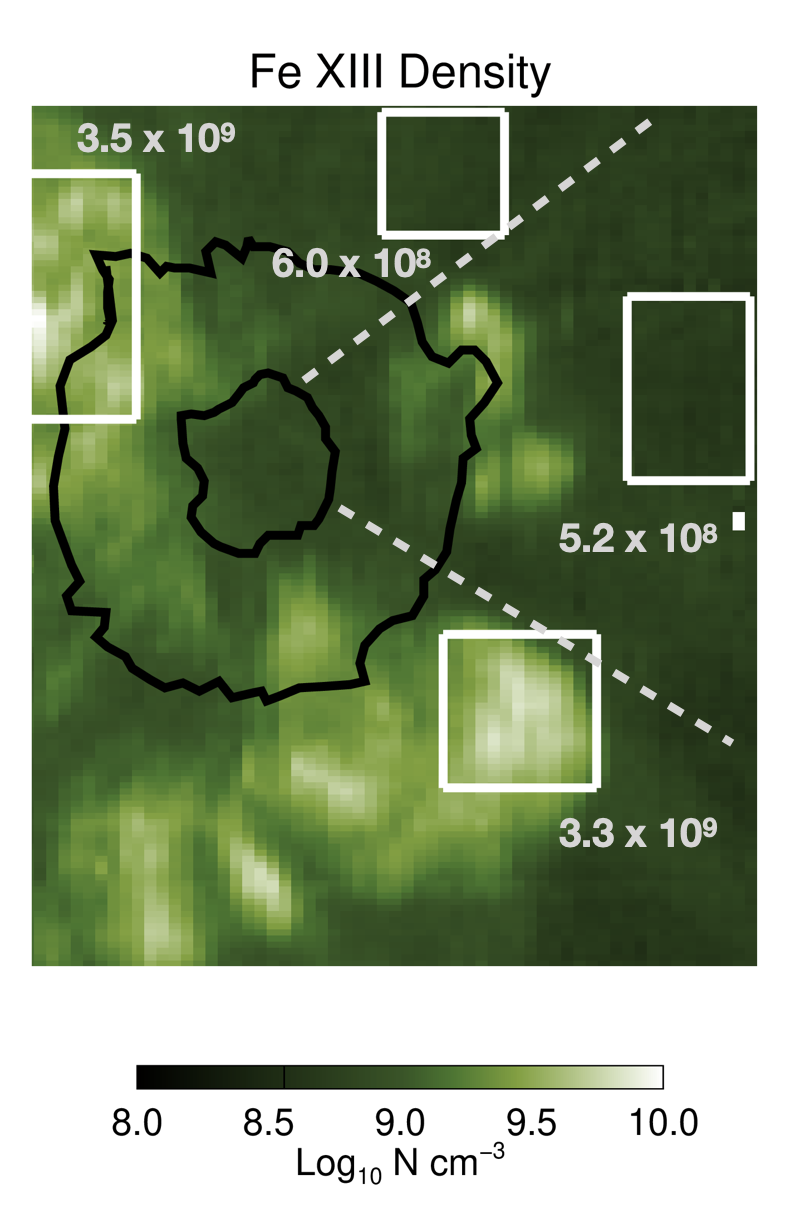}
\includegraphics[width=0.18\textwidth]{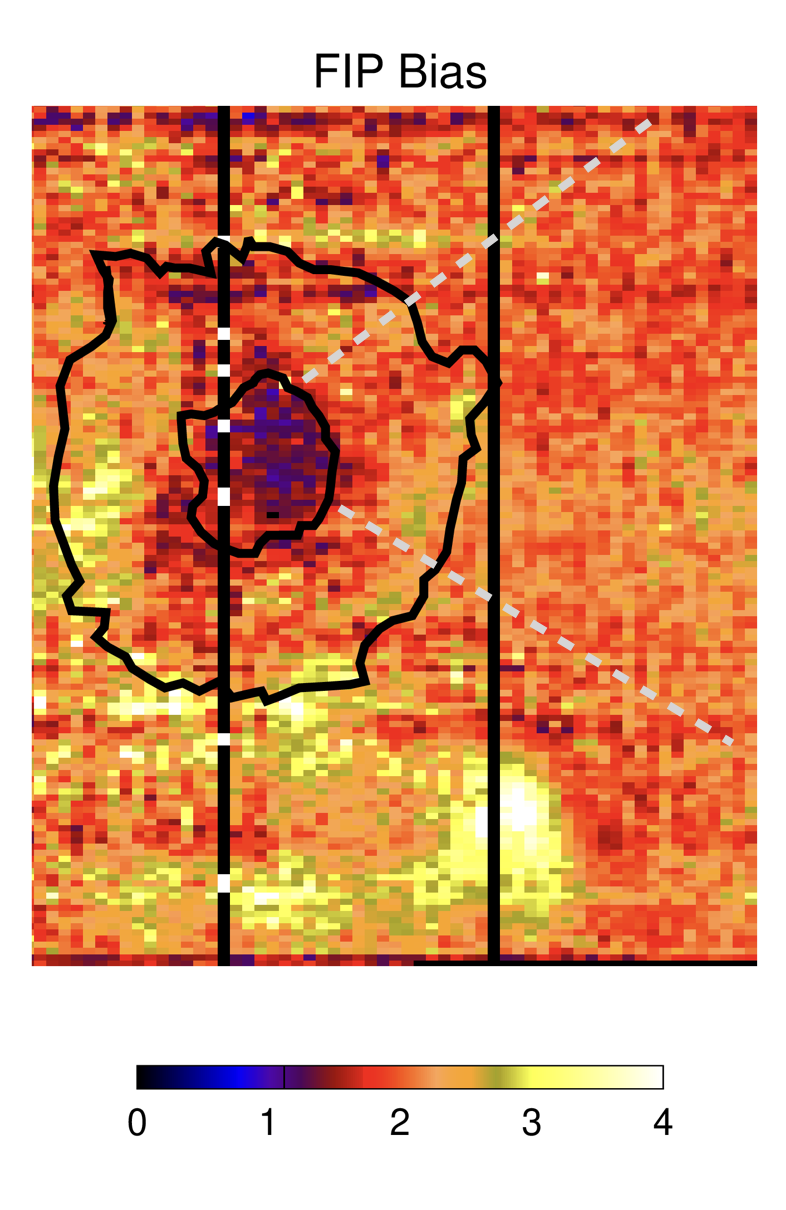}

\caption{\emph{Hinode}/EIS \ion{Fe} {12} intensity, Doppler velocity, non-thermal velocity maps, \ion{Fe} {13} density and \ion{Si}{10}/\ion{S}{10} FIP bias (or composition) maps at 07:24 UT on 20 May 2016. \emph{SDO}/HMI continuum contours show the locations of the umbra and penumbra boundary.  Dashed lines enclose the fan loops rooted on the western side of the sunspot.  The EIS FOV is X = [$-$60$\arcsec$, 60$\arcsec$] and Y = [$-$170$\arcsec$, $-$30$\arcsec$] (coordinates from Sun center). Regions in the density map sample loop populations with mean density values provided for each white box.   (The black stripes in the composition map are artefacts of the CCD and should not be confused with the IBIS FOV boundaries shown in the left panel of Figure \ref{fig_fip}).}
\label{fig_eis}
\end{figure*}

\subsection{\emph{Hinode}/EIS Observations and Methods}
 \emph{Hinode} EIS spectral data were used for the plasma composition analysis of the sunspot within AR 12546.
A FOV of 120$\arcsec$$\times$160$\arcsec$ was created using the 2$\arcsec$ slit in 2$\arcsec$ steps, taking 60 sec exposures at each slit position.
The single scan began at 07:24 UT on May 20 and finished two hours later.
Study $\#$404 (Atlas$\_$60) is a full spectral atlas of both CCDs therefore it contains the diagnostic spectral lines required for constructing a spatially resolved composition map.

 Data reduction was carried out using the eis$\_$prep routine that is available in Solar SoftWare \citep{freeland98}.
The CCD signal in each pixel was converted into calibrated intensity units of erg cm$^{-2}$ s$^{-1}$ sr$^{-1}$ $\angstrom$$^{-1}$ and pixels affected by cosmic ray hits, dust, and electric charge were removed/replaced.
All data were corrected for instrumental effects of orbital spectrum drift \citep{kamio10}, CCD spatial offsets, and the grating tilt.

\begin{table}[]
    \centering
    \begin{tabular}{lcc}
    \hline
Diagnostics     & Ion &Wavelength ($\angstrom$)  \\
    \hline        
    Emission Measure & \ion{Fe}{8} & 185.21, 186.60\\     
    & \ion{Fe}{9} & 188.50, 197.86\\
    & \ion{Fe}{10} & 184.54\\
    & \ion{Fe}{11} & 188.21(b)\\
    & \ion{Fe}{12} & 192.39, 195.12(b)\\
    & \ion{Fe}{13} & 202.04, 203.83(b)\\
    & \ion{Fe}{14} & 264.79, 270.52(b)\\
    & \ion{Fe}{15} & 284.16\\
    & \ion{Fe}{16} & 262.98\\
    & \ion{Fe}{17} & 254.87\\
     \hline
    FIP Bias & \ion{Si}{10} & 258.38\\
     & \ion{S}{10} & 264.23\\
      \hline
    Density & \ion{Fe}{13} &202.04, 203.83(b)\\
    \hline
    \end{tabular}
    \caption{\emph{Hinode}/EIS emission lines used for various plasma diagnostics in this study. Blended lines (b) are fit with multiple Gaussian functions. }
    \label{t:EIS_lines}
\end{table}

To construct the composition map, spectral lines from consecutive ionization stages of \ion{Fe}{8}--\ion{Fe}{17} and the low FIP \ion{Si}{10} (FIP = 8.15 eV) and high FIP \ion{S}{10} (FIP = 10.36 eV) were fit with single Gaussian functions except where the lines are blended in which case the line was fit with multiple Guassian functions.
The \ion{Si}{10}/\ion{S}{10} line ratio was used to determine FIP bias and the density was measured with the \ion{Fe}{13} 202.04 $\angstrom$/203.83 $\angstrom$ line ratio. 
The specific emission lines are given in Table \ref{t:EIS_lines}.
The CHIANTI Atomic Database, Version 8.0 \citep{dere97,delzanna15} was used to carry out the contribution function calculations, applying the photospheric abundances of \cite{grevesse07} for all of the spectral lines while assuming the measured \ion{Fe}{13} densities.
The Markov-Chain Monte Carlo (MCMC) algorithm contained within the PINTofALE software package \citep{kashyap00} was used to compute the emission measure (EM) distribution for the Fe lines.
The EM distribution was then convolved with the contribution functions and fit to the observed intensities of the low-FIP Fe spectral lines.
Si is also a low-FIP element therefore the EM derived from the Fe lines was scaled to reproduce the intensity of the \ion{Si}{10} line.
Finally, the FIP bias was determined to be the ratio of the predicted to observed intensity for the high FIP \ion{S}{10} line. 
The estimated uncertainty of the FIP bias ratio is 0.30 assuming an intensity error of 20$\%$.
A full account of the method is available in \cite{brooks15} and \cite{baker18}.

\ion{Fe}{12} 195.12 $\angstrom$ relative Doppler velocities were measured versus a reference wavelength defined by averaging the centroid wavelengths of all pixels within the data array.
This method was adopted as EIS does not have an absolute wavelength calibration.
Excess broadening in the deblended \ion{Fe}{12} 195.12 $\angstrom$ emission line spectra was calculated from 
\begin{equation}\label{eq_ntv}
\delta\lambda = \frac{\lambda_{0}}{c}~\sqrt{4~ln~2 \left(\frac{2~k_{B}T_{i}}{m}~+~\xi^{2} \right)~+~\sigma_{I}^{2}} ,
\end{equation}
where $\delta$$\lambda$ is the observed line width, $\lambda$$_{0}$ is the line centroid, k$_{B}$ is Boltzmann's constant, T$_{i}$ is the ion temperature, $m$ is the mass, $\xi$ is the nonthermal velocity, and $\sigma$$_{I}$ is the instrumental width \citep[e.g.][]{brooks16}.
Figure \ref{fig_eis} shows the \emph{Hinode}/EIS \ion{Fe}{12} intensity, Doppler and nonthermal velocity maps, \ion{Si}{10}/\ion{S}{10} composition map, and \ion{Fe}{13} density map at 07:24 UT on 2016 May 20.
\emph{SDO}/HMI continuum contours have been overplotted on the EIS maps to mark the umbral and penumbral boundaries of the sunspot.

Above the umbra, the plasma is blue-shifted with upflow speeds of 10--20 km s$^{-1}$ and nonthermal velocities range from $\sim$15 km s$^{-1}$ above the center to $\sim$30 km s$^{-1}$ toward the umbra/penumbra boundary.  Plasma density is $\sim$5$\times$10$^{8}$ cm$^{-3}$.
Plasma composition above the umbra is photospheric with a FIP bias of 1 above its core; toward the boundary, the composition becomes more fractionated with FIP bias $\sim$1.5--2, especially on the eastern edge.

Plasma parameters are more extreme above the penumbra where the loops surrounding the sunspot are rooted.
Upflows transition to downflows in the loops at the outer boundary.
Nonthermal velocities are 30--50 km s$^{-1}$ in loops located to the east and south of the penumbra; they are $\sim$30--40 km s$^{-1}$ in the west (in between the dashed lines in the EIS maps).
Plasma density increases by an order of magnitude above the penumbra compared with the umbra.
FIP bias exceeds 3$^{+}$ above the eastern penumbra and reaches 4$^{+}$ at the boundary in the east and to the south.
The strongest fractionation is located in the southern region in the vicinity of the highest nonthermal velocities of 45--50 km s$^{-1}$.

 \section{Coronal Loop Connectivities}
\begin{figure}[ht]
    \centering
     \includegraphics[width=0.48\textwidth]{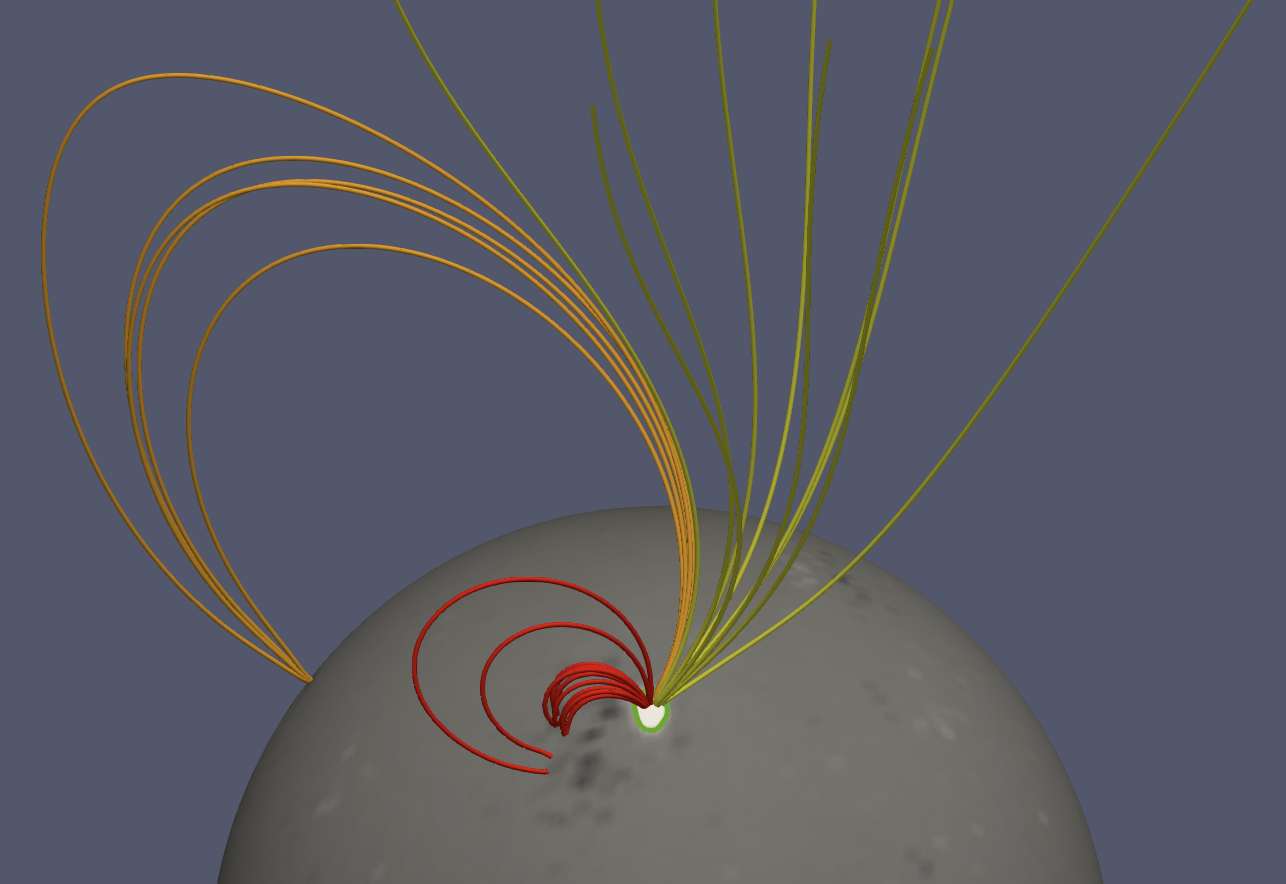}
 \caption{Selected field lines of the PFSS extrapolation of AR 12546 based on an \emph{SDO}/HMI synoptic radial field magnetogram at 13:00 UT on 2016 May 20.  Field lines are color-coded as closed within the same active region (red), closed to a neighboring active region (orange), or open (yellow); the radial magnetic field saturated at 100 G is represented in greyscale, and the green contour on the sunspot represents its 500 G isocontour. }
    \label{fig_pfss}
\end{figure}

A Potential Field Source Surface (PFSS) extrapolation was computed to model the coronal loop system surrounding the sunspot.
The rationale behind the use of the PFSS extrapolation is that this model captures the global field of the sunspot for comparison with the \emph{SDO}/AIA coronal images of Figure \ref{fig_context}.
The PFSSPY package \citep{pfsspy_yeates,pfsspy_stansby} was employed to extrapolate the coronal field from the HMI synoptic radial field map of CR2177.
Figure \ref{fig_pfss} shows the extrapolation with selected field lines. 
A green contour represents 500 G in the positive polarity sunspot.
In general, there is good qualitative correspondence of the loops in the \emph{SDO}/AIA images at the times of the EIS and IBIS observations with field lines in the extrapolation (Figure \ref{fig_pfss}).

The coronal loop configuration of AR 12546 is characteristic of a bipolar region with distinct loop populations.
Yellow field lines on the western side of the sunspot are long, extended loops that reach the source surface of the PFSS model (=2.5 R$_{sun}$) therefore these field lines are considered to be open. 
To the north, the orange, long loops are connected with the negative polarity of the active region located to the north-east of AR 12546.
The plasma density in these regions is $\sim$5--6$\times$10$^{8}$ cm$^{-3}$ (see the mean densities within the boxes of the density map in Figure \ref{fig_eis}) and plasma composition is partially fractionated with FIP bias of 1.5--2.
In contrast, the red loops on the east and south of the sunspot are compact loops that connect mainly with the opposite polarity within the active region.
The density of the short, closed loops is an order of magnitude higher and the plasma is highly fractionated with FIP bias of 3--4.

\section{Alfv\'enic Perturbations in the Sunspot Chromosphere}\label{sec_alfven}

\begin{figure*}[ht]
\centering
\includegraphics[width =0.98\textwidth]{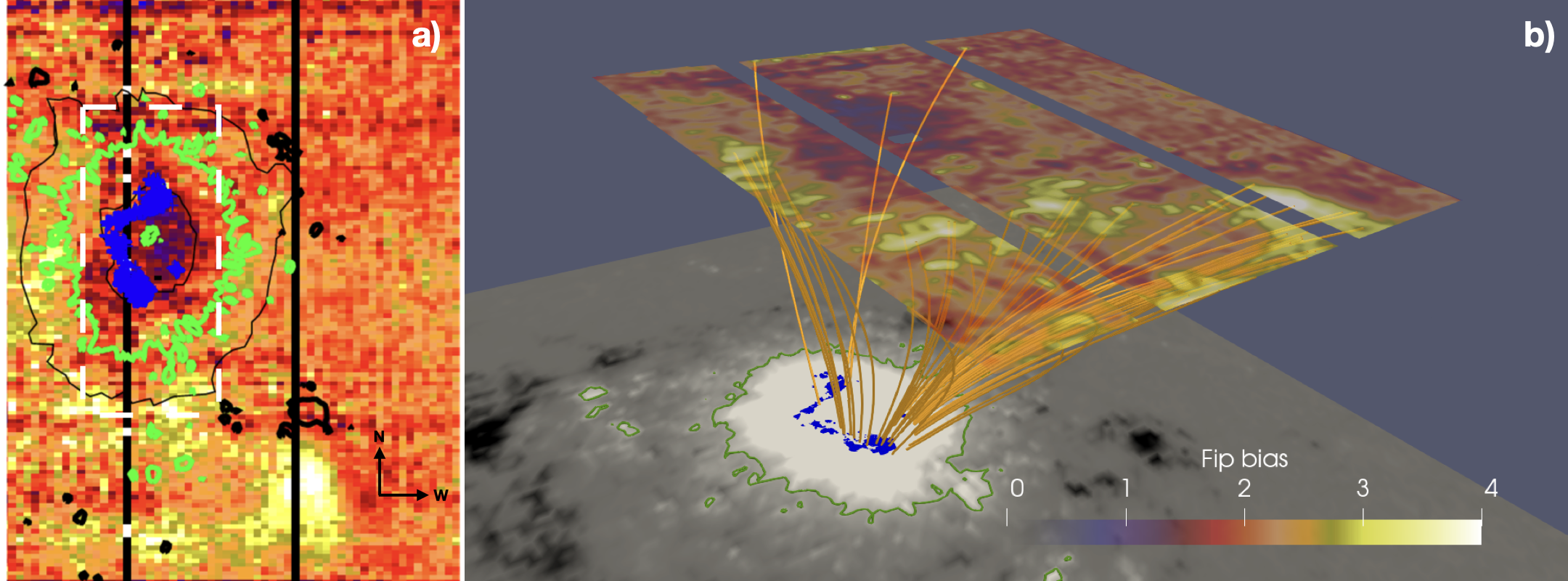}
\caption{(a)  \emph{Hinode}/EIS \ion{Si}{10}/\ion{S}{10} FIP bias map at 07:24 UT on 2016 May 20. Contours:  (i)  \emph{SDO}/HMI continuum umbra and penumbra boundaries  (thin black lines), (ii) \emph{SDO}/HMI LoS magnetogram $\pm$500 G (thick green/black lines = positive/negative polarity), (iii) Location of Alfv\'enic waves from IBIS \ion{Ca}{2} observation (blue dots),
(iv) IBIS FOV (white dashed box).
(b) Selected high-FIP bias field lines connecting high FIP bias in the corona to the umbra-penumbra boundary (orange) started from values of FIP bias $>$ 2.7. The height of the FIP bias map is $z=$3.6 CEA-deg and the value of the linear force-free parameter $\alpha=-0.2$  CEA-deg$^{-1}$ (The conversion factor is 1 CEA-deg $\simeq$ 12.17 Mm).
The green contour represents the 500 G isocontour of the vertical magnetic field in the SHARP data.
}
\label{fig_fip}
\end{figure*}
\cite{stangalini20} found the presence of Alfv\'enic perturbations in the sunspot chromosphere in the 3-minute band of the IBIS \ion{Ca}{2} time series (see Section \ref{sec_ibis}). The frequency band corresponding to this period, which is also the dominant period in the solar chromosphere \citep[e.g.][and references therein]{khomenko15}, is smaller than the ion-cyclotron frequency, thus in the regime appropriate for the model of \citet{laming15}.
A phase lag analysis between $CP$ and intensity ruled out the possibility of opacity and other spurious effects by identifying a specific phase lag of the order of $-35$ degrees between the two quantities, which is not consistent with the phase values expected from cross-talk or radiative opacity effects. 
Due to the vertical gradient of the magnetic field, any plasma density perturbation can induce a height variation of the response function of the spectral line, thus resulting in an observed spurious magnetic oscillation which is merely a consequence of an opacity change and has nothing to do with a real magnetic wave. In this regard, the study of the phase relation between different diagnostics is helpful in the identification of real magnetic oscillations. 
Indeed, by collecting phase measurements corresponding to high coherence it was possible to discriminate between different effects and identify real magnetic oscillations in the sunspot chromosphere. Coherence is independent of the wave amplitude, thus this technique is able to detect correlations between two signals even if their amplitudes are small.
It is worth noting that Alfv\'enic shocks were independently detected by \cite{houston20} at the same spatial locations, thereby confirming the interpretation of these disturbances in terms of real magnetic fluctuations.
The locations of the Alfv\'enic perturbations are indicated by the blue dots overplotted on the \emph{Hinode}/EIS FIP bias map of Figure \ref{fig_fip}(a) and on the $CP$ map in Figure \ref{fig_ibis}(d).
The dots are aligned in a distinct C-shaped structure running from the north to the south along the eastern edge of the sunspot umbra.


Within the same EIS FOV in Figure~\ref{fig_fip}(a), coronal loops containing highly fractionated plasma are present with particularly high values both to the east and to the south-west of the sunspot.
The projected spatial proximity of the C-shaped structure and high FIP bias values poses the question whether the loops of highly fractionated plasma observed in the corona are magnetically connected to the specific locations of Alfv\'enic perturbations in the chromosphere identified by \cite{stangalini20}.
The context to this question is provided by theoretical models based on the hypothesis that the fractionation process producing the FIP-effect is powered by the conversion of magnetic waves at chromospheric heights \citep[e.g.][]{schwadron99,laming15}.

In order to answer this question, a magnetic model of the sunspot area is needed that is representative of the sunspot field so that we can determine if the field lines threading regions of high FIP bias values are rooted in the blue dots of Figure \ref{fig_fip}. 
The main difficulty is that the FIP bias map is the result of a pixel-dependent, line-integrated emission of coronal lines to which it is challenging to attribute a height, and even more so a single height for the entire map.
In addition, the paths followed by the modelled field lines depend on the properties of the chosen magnetic field model.  
Finally, there are more than eight hours between the beginning of the EIS observation, which was used to compute the FIP bias map, and the end of the IBIS observation, which was used to identify the Alfv\'enic perturbations. 
This complicates the alignment of the different observations, especially considering that both EIS and IBIS lack absolute pointing information.
Therefore, the choice of the time of the magnetic field observations that are used to build the magnetic field model is also a factor of uncertainty.  
Given these difficulties, we adopted a heuristic approach by testing whether a combination of magnetic model and height of the FIP bias map exists where field lines starting from areas of high FIP bias values are rooted in the proximity of the blue dots.

A magnetic model for the sunspot can be obtained using a force-free extrapolation of photospheric measurements \citep[see, e.g.][]{wiegelmann12}.
The photospheric magnetogram used as input for the extrapolation is the \emph{SDO}/HMI SHARP magnetogram taken at 13:00 UT, in between the end of EIS rastering and the start of the IBIS observation.
SHARP data provide vector magnetograms of regions of the Sun in a Cylindrical Equal Area (CEA) projection, where the spatial dimensions are given in CEA degrees with 0.03 CEA-deg $\simeq$ 0.5$\arcsec \simeq 365$ km  at the center of the disk.
We use this approximate conversion factor in the scale estimations given below.
The full SHARP field presents a significant flux imbalance (about 15\%), therefore we use the linear force-free extrapolation method of \cite{Seehafer1978} which does not require strict flux balance to be enforced.
The entire FOV of the SHARP magnetogram covering an area of 22.89$\times$12.99 CEA-deg was used to compute the Fourier coefficient of Seehafer's solution.
However, in order to reduce the computing time required by the parametric study described below, the extrapolated magnetic field was computed in a smaller volume above the sunspot of size [11.22, 11.22, 16] CEA-deg. 
The extrapolation volume was discretized with a uniform resolution of 0.06 CEA-deg in the horizontal directions and with an exponentially stretched axis in the vertical direction such that the pixel size increases from 0.06 at the bottom to 0.96 at the top. 

The ratio of the vertical current density to the vertical magnetic field,  $\alpha_{NL}=J_z/B_z$, represents the local torsion of field lines and is constant along individual field lines in force-free extrapolations. Within the linear approximation, this function is a free, constant parameter that is limited in magnitude by the inverse of the linear dimension of the extrapolated magnetogram.
For Seehafer's method applied to the entire FOV of the SHARP magnetogram, $\alpha_{\rm max}=$ 0.27 CEA-deg$^{-1}$ (corresponding to 0.022 Mm$^{-1}$).
Different magnetic field models are then obtained for different values of the constant $\alpha$ ($|\alpha| < \alpha_{\rm max}$) and the same magnetogram at the bottom boundary.

The alignment problem between the magnetic model based on the reprojected SDO/SHARP magnetogram on the one hand, and the plane-of-the-sky SDO/AIA, IBIS, and EIS observations on the other was treated as follows.
First, the locations of the Alfv\'enic perturbations were derotated from 13:39 UT (starting time of the IBIS measurements) to the starting time of the EIS raster.
Second, since both EIS and IBIS lack absolute pointing information, the EIS raster and derotated locations of Alfv\'enic perturbations were aligned to multiple images of SDO at the time of the EIS raster. 
In particular, this operation produces a co-alignment between the (derotated) location of the Alfv\'enic perturbations and the HMI line of sight magnetogram at 07:24~UT.
Finally, a 500 G isocountour of the latter was used to match a similar contour in the SHARP vertical magnetic field (at 13:00~UT) to co-align the location of the  Alfv\'enic perturbations and FIP bias map with the extrapolated field of the coronal models.

\begin{figure*}[ht]
\centering
\includegraphics[width=0.95\textwidth]{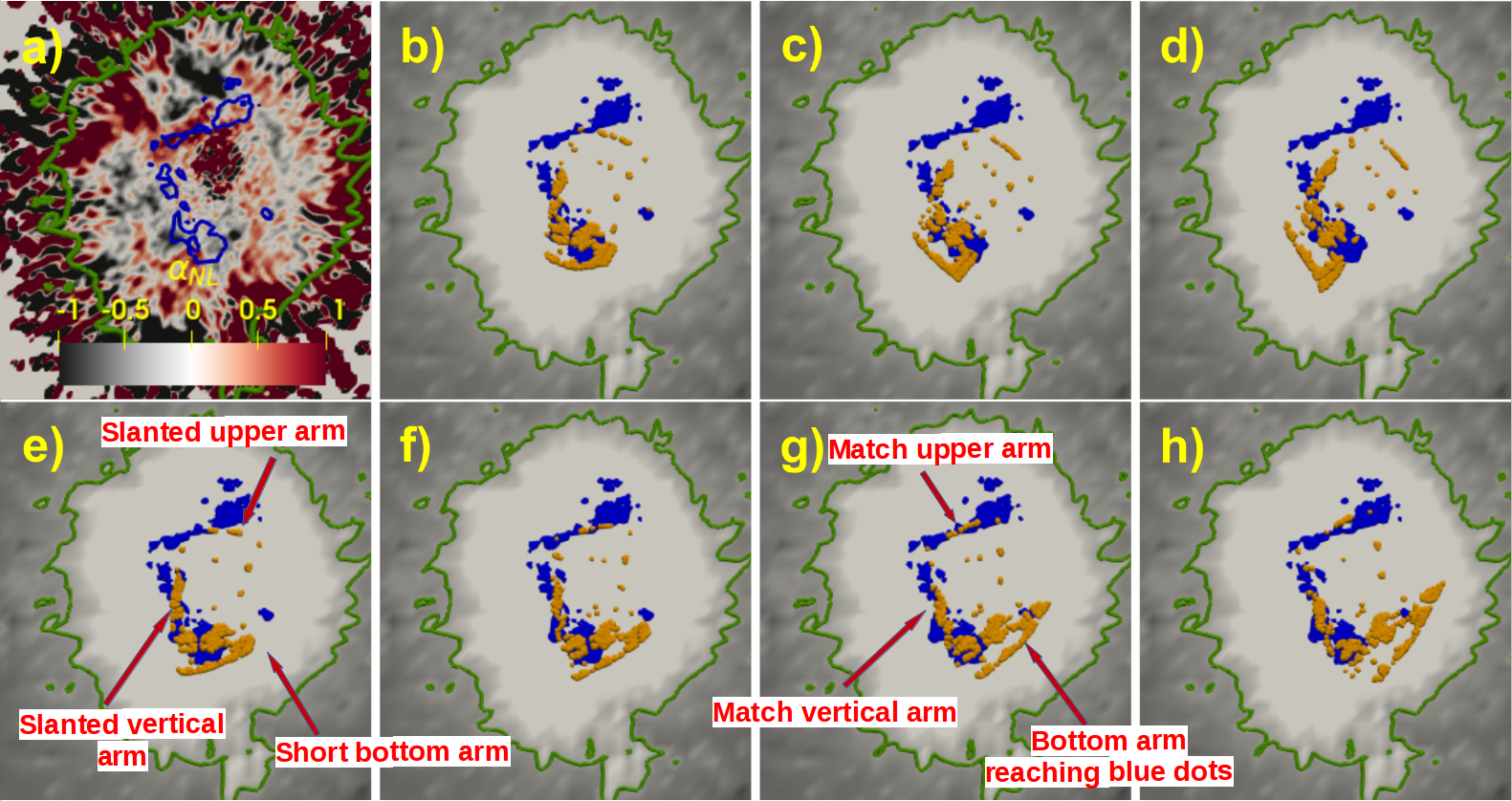}
\caption{(a) Distribution of $\alpha_{NL}$ on the sunspot, saturated between $-$1 and 1, derived from the SHARP magnetogram. The blue contour identifies the spatial distribution of the blue dots in panels (b-h).
The remaining panels (b-h) illustrate the match between the footpoints of high FIP bias field lines (orange dots) and the location the Alfv\'enic perturbation (blue dots), for positive values of $\alpha=$0.1, 0.2, 0.25 (panels b-d, respectively), for $\alpha=0$ (panel e), and for negative values of $\alpha=-0.1, -0.2, -0.25$ (panel f-h, respectively), in CEA-deg$^{-1}$.
The selected heights of the FIP bias maps are 3.6, 4.2, 4.5, 3.3, 3.3, 3.6, 3.6 CEA-degrees for cases (b-h), respectively. 
The conversion factor is 1 CEA-deg $\simeq$12.17 Mm. 
On all panels, the green contour represents the 500 G isocontour of the vertical magnetic field in the SHARP data.
Annotations on panels e) and g) highlight the matching criteria discussed in the text.}
\label{fig_dots_match}
\end{figure*}

In summary, the heuristic method that we adopted consists of the following steps: 
\begin{enumerate}
\itemsep-5pt
    \item producing magnetic field models for different (constant) $\alpha$ values between $-\alpha_{\rm max}$ and $+\alpha_{\rm max}$; 
    \item for each field model, placing the co-aligned FIP bias map at different heights; 
    \item for each height of the FIP bias map, tracing the field lines in the given model starting from FIP bias values above 2.7 in the entire EIS FOV. 
    The value 2.7 is chosen as the lowest possible value consistent with capturing most of the yellow area in the Figure~\ref{fig_fip}(a), but reducing the number of pixels at the edge of the FIP bias map. This is done to avoid that the rectangular shape of the EIS FOV produces a misleading boundary of the field line distribution in the following steps. 
    The filtering of field line seeds results in selecting relatively high-FIP bias field lines. As an example, Figure~\ref{fig_fip}(b) shows the 3D rendering of the spatial arrangement of Alfv\'enic oscillations (blue dots), a few selected field lines (orange), and the EIS map for the parameters of Figure~\ref{fig_dots_match}(g);
    \item flagging the footpoints of the field lines with high FIP bias at chromospheric heights (1$\arcsec$); 
    \item comparing the location of the footpoints of field lines with high FIP bias with the location of the co-aligned Alfv\'enic oscillations; 
    \item finally, verifying if combinations of $\alpha$ and FIP bias map height exist that produce a distribution of high FIP bias field line footpoints similar to the distribution of blue dots in Figure~\ref{fig_fip}(a). 
\end{enumerate}
The above procedure yielded the maps in Figure~\ref{fig_dots_match}(b-h), where, in addition to the blue dots representing the location of  Alfv\'enic oscillations, the footpoints of the high FIP bias field lines are shown as orange dots. 
These maps answer our initial question: there are indeed combinations of model parameters for which the orange and blue dots are closely located. 
In other words, our parametric, heuristic study supports the existence of a magnetic link between high FIP bias values and the locations of Alfv\'enic perturbations in the chromosphere identified by \cite{stangalini20}.

One can further try to deduce which is the value of $\alpha$ that results in the best match between the orange and blue dots in Figure~\ref{fig_dots_match}(b-h). 
This is inevitably very subjective, as it strongly depends on which subset of dots is given priority in the match.
If, for instance, only the number of orange dots overlapping the blue structure is considered, one would likely choose the potential case ($\alpha=0$ in panel e) or even a slightly positive value of $\alpha$ as the best matching case. 
This criterion discards the more isolated dots in the center and upper parts of the sunspot as not significant.
On the other hand, if the shape of the distribution is chosen as the primary matching criteria, then one may recognize how, similarly to the blue dots, the orange dots in Figure~\ref{fig_dots_match}(b-h) are arranged in a pattern roughly shaped as a C with straight arms (except for few orange dots in the center of the sunspot).
The red arrows on panels (e) and (g) of Figure~\ref{fig_dots_match} indicate how the arms of the C-shape of the orange dots in two cases are identified.
For instance, as the annotations in panel (e) indicate, in the potential case both the vertical and upper orange arms are at an angle with the corresponding blue arms, whereas we deem the overlap of the whole structures to be better in panel g. 
By matching the blue and orange C-shapes as a whole, we would then identify the $\alpha=-0.2$ CEA-deg$^{-1}$ as best matching case. 

Hence, adopting different matching criteria results in different values for the best matching $\alpha$, pointing at the limitations that a constant-$\alpha$ model of the magnetic field has in this particular application.
Also, we stress again that to treat the FIP bias map as a flat horizontal plane at a given height is a very crude approximation. 
On the other hand, the range of heights between 40 and 55 Mm of the FIP bias maps identified in Figure~\ref{fig_dots_match}(b-h) yields coronal electron temperatures in the range of $\approx$ [0.9, 1.2] MK, assuming hydrostatic equilibrium  \citep{aschwanden02}.
This is consistent with the temperature range for which the \ion{Si}{10}--\ion{S}{10} FIP bias line-pair is an effective diagnostic of coronal plasma composition \citep{feldman09,brooks11a}

The space-dependent distribution of $\alpha$ in the sunspot area can be computed as $J_{\rm z}/B_{\rm z}$ using the photospheric observation of the SHARP magnetogram.
This is plotted in Figure~\ref{fig_dots_match}a, with the blue contour identifying the distribution of the blue dots, showing that, while still  predominantly positive, $\alpha$ values of both signs are present. 
It is interesting to note that a C-shaped concentration of negative $\alpha$ values is present right to the east of the C-shaped contour of the blue dots. 
The blue contour and the negative $\alpha$ values are not exactly overlapping, but their appearance is very similar in shape with just a slight shift (of about 0.3 CEA-deg) that is comparable to the alignment accuracy among the different instruments/maps. 
The possible spatial correlation between Alfv\'enic oscillations and negative values of $\alpha$ suggested by Figure~\ref{fig_dots_match}(a) is interesting because it may give some clues about the background field in which the oscillations took place.
However, given the limitations of alignment and modeling discussed so far, we do not pursue this point any further, and leave the investigation of this aspect to future studies on the nature of the Alfv\'enic oscillations (e.g. Section \ref{concl}).

In summary, the heuristic method defined two parameters (the magnetic twist and the height of the FIP bias map) that resulted in a positive spatial correlation between high values of FIP bias and the Alfv\'enic 3-minutes magnetic oscillations detected in the IBIS \ion{Ca}{2} time series by \cite{stangalini20}. 
This is indicative of a magnetic connection between chromospheric magnetic oscillation regions and high values of FIP bias observed in the corona above the sunspot.
Given the intrinsic limitations of observations and magnetic modeling discussed above, these findings are nevertheless consistent with the theory of FIP fractionation outlined in \cite{laming15}.

\section{Discussion}
 In this work, we have investigated the spatial distribution of coronal plasma composition in the vicinity of the large sunspot of AR 12546.
 \emph{Hinode}/EIS observations revealed that compositional variation ranged from little or no fractionation in the corona above the core of the umbra to partially or highly fractionated plasma in groups of coronal loops rooted in the surrounding penumbra.
 
 Using a PFSS extrapolation of the large scale magnetic field of the positive-polarity sunspot, we identified distinct loop populations based on their connectivities:  open field or long loops connected externally to the negative polarity of an active region $\sim$ 570 Mm to the northeast and short, dense loops internally connected to the negative polarity of the active region.
 
 Plasma composition is relatively uniform within each loop population but varies across populations.
 FIP bias was 1.5--2 in the open field or long loops connecting externally to a another active region whereas it was 3--4 in the loops connecting to the opposite polarity within AR 12546 (Figure \ref{fig_eis}).
 Loops within a particular group are likely to share similar global properties and evolve together.
 In AR 12546, the very simple bipolar topology coupled with little evolution of the large-scale magnetic field prior to the EIS and IBIS observations, suggest that there was limited mixing of plasma compositions via magnetic reconnection of different loop populations.
 
 Among the global properties shared by coronal loops within the same population, loop length and resonance frequencies are at least theoretically related to the observed plasma fractionation distribution in and around the sunspot.
The Laming ponderomotive force fractionation model predicts that resonant Alfv\'enic waves, generated by nanoflaring reconnection in the corona, increase plasma fractionation in the vicinity of footpoints of resonant closed loops whereas there is little resonance and fractionation along open field. 
In support of the Laming model, numerical simulations of \cite{dahlburg16} showed that ponderomotive acceleration occurs at loop footpoints as a consequence of MHD waves generated by magnetic reconnection in the corona and that the FIP effect is a natural outcome of coronal heating. 
Significantly, the ponderomotive acceleration increases with increasing temperature and with decreasing length in closed coronal loops in the simulations.
In the corona above the umbra of the large sunspot, the field had photospheric plasma composition.
At photospheric and chromospheric heights, sunspot umbrae are regions of temperature minima so that elements are mainly neutral or singly ionized \citep[e.g.][]{loukitcheva14,lodders19}.
As a consequence, local temperatures are not high enough to create a sufficient reservoir of ionized elements in the chromosphere, inhibiting plasma fractionation. 
Plasma transported from the chromosphere to the coronal field above the umbra is therefore likely to be unfractionated photospheric plasma.

 On the western side of the sunspot, which is magnetically connected to another active region or has open field lines, FIP bias values were somewhat higher,  between 1.5 and 2. 
 The highest FIP bias of 3--4 was found on the eastern/southern sides of the spot in high temperature active region core loops. 
 These loops are rooted in the penumbra and are subject to convection-driven footpoint motions leading to higher frequency heating than in the loops rooted in the umbra \citep[see][and references therein]{gdz18}.

 Our results are consistent with the predictions of the Laming model and the output from the simulations of \cite{dahlburg16} in that high FIP bias plasma occurs in high temperature, short loops where nanoflaring in the model generates high Alfv\'enic wave flux which is amplified by resonance.
 In the other sets of loops with different connectivities e.g., open field lines, lower wave flux is expected as there is no repeated reflection of any potential Alfv\'enic waves created there, and consequently, these loops are likely to contain plasma with lower FIP bias. 
 In fact, these findings do not exclude other theoretical models based on either wave interactions with chromospheric ions e.g. the ion cyclotron wave heating model of \cite{schwadron99} or on processes linked to coronal heating  e.g. heat conduction of the thermoelectric driving model of \cite{antiochos94}. 
 
 In the wider context, we still fundamentally lack an understanding of what is happening in the chromosphere when we see activity in the corona, and a key goal generally is to elucidate the connection between activity in the low atmosphere and observable changes higher up. We have found that the internally connecting core loops with the highest FIP bias are rooted in areas where the Alfv\'enic perturbations were found in the chromosphere.
This is the first observational evidence of detectable Alfv\'enic perturbations in the chromosphere being linked to coronal loops containing highly fractionated plasma. 
Whether this is the result of a response to coronal heating, as in the ponderomotive force model, or further evidence of heating at coronal heights being driven from below, remains an open question. 

\section{Concluding Remarks}\label{concl}
This work represents a first attempt to investigate the role of magnetic fluctuations in plasma fractionation, made possible thanks to nearly simultaneous observations at chromospheric and coronal heights by IBIS and EIS, respectively. 
Our results demonstrate a possible link between magnetic perturbations observed at chromospheric heights as small fluctuations of the spectropolarimetric quantities, and the locations of high FIP bias observed in the corona. 
They therefore observationally support a role for MHD waves in the generation of the FIP effect and wave-based theoretical models.

Some questions still remain open and identify possible future research directions.
As already mentioned, magnetic fluctuations are expected for a number of different MHD wave modes. 
These can be magneto-acoustic modes which are locally excited within the umbra by residual convection and/or p-mode absorption, or globally excited eigenmodes of the sunspot. 
In this study, the exact identification of the wave mode responsible for the observed magnetic fluctuations was not possible with the available data. 
Indeed, different modes can co-exist in the same structure hampering the identification process.
However, as noted in \cite{stangalini20}, the locations where the Alfv\'enic waves are observed correspond to a narrow range of magnetic field inclinations, suggesting a possible role of the magnetic field geometry.
In this regard, it is worth recalling that MHD waves in magnetic structure can undergo a mode conversion at the Alfv{\'{e}}n-acoustic equipartition layer, with part of the energy contained in the acoustic-like components (fast MHD mode in the plasma-$\beta > 1$ regime) being converted to a combination of fast magneto-acoustic (in the plasma-$\beta < 1$ regime) and magnetic-like waves. 
This physical mechanism is dependent on the attack angle between the wavevector and the field lines \citep[see for instance][]{gary01,newington10, cally11, hansen16, cally11}, thus \cite{stangalini20} speculated on the possible role of the mode conversion and magnetic field geometry in the appearance of magnetic waves at chromospheric heights.
In support of this scenario, \cite{grant18} uncovered evidence for Alfv\'enic waves, with the observed signatures being consistent with induced ponderomotive forces at the umbra/penumbra boundary of a sunspot chromosphere, suggesting that such wave-coupling effects may be linked to the increasing attack angles found in these locations. 

Nevertheless, the Alfv\'enic waves identified by \cite{stangalini20} might be associated with different wave modes not all necessarily producing the fractionation, however, they could be considered a proxy to identify the spatial locations where, given a magnetic field geometry, the conversion of acoustic-like to magnetic-like waves is particularly efficient.

In addition to the unambiguous identification of the wave process responsible for the magnetic waves, another important aspect is the propagation direction of the MHD waves, which plays a significant role in the FIP and I-FIP model of \cite{laming15}. 
Both the equipartition layer and the transition region can represent a reflective mirror for different types of waves \citep[e.g.][]{hansen16}. 
For this reason, MHD waves can undergo several reflections/conversions, thus leaving open both the possibilities of waves coming from below or above (i.e. due to nanoflares).

In our view, these remain important aspects to be further addressed in the future, and they may provide useful information to constrain and validate existing theoretical models.
The heuristic method employed in this study combines the linear extrapolation of photospheric magnetic field and the FIP bias map to investigate the possible connectivity of the Alfv\'enic chromospheric perturbations identified by \cite{stangalini20}. 
However, further studies that involve the full inversion of the chromospheric spectropolarimetric signals at the location of the perturbations are required to properly identify the properties of the background and transient magnetic field. 

\acknowledgments
Hinode is a Japanese mission developed and launched by ISAS/JAXA, collaborating with NAOJ as a domestic partner, and NASA and STFC (UK) as international partners. 
Scientific operation of Hinode is performed by the Hinode science team organized at ISAS/JAXA. 
This team mainly consists of scientists from institutes in the partner countries. 
Support for the post-launch operation is provided by JAXA and NAOJ (Japan), STFC (UK), NASA, ESA, and NSC (Norway). 
\emph{SDO} data were obtained courtesy of NASA/\emph{SDO} and the AIA and HMI science teams.
D.B. and D.S. are funded under STFC consolidated grant number ST/S000240/1 and L.v.D.G. is partially funded under the same grant.
G.V. acknowledges the support  from the European Union's Horizon 2020 research and innovation programme under grant agreement No 824135 and of the STFC grant number ST/T000317/1.
The work of D.H.B. was performed under contract to the Naval Research Laboratory and was funded by the NASA Hinode program.
D.B.J. wishes to thank Invest NI and Randox Laboratories Ltd for the award of a Research $\&$ Development Grant (059RDEN-1).
S.J. acknowledges support from the European Research Council under the European Unions Horizon 2020 research and innovation program (grant agreement No. 682462) and from the Research Council of Norway through its Centres of Excellence scheme (project No. 262622).
The authors wish to acknowledge scientific discussions with the Waves in the Lower Solar Atmosphere (WaLSA; www.WaLSA.team) team, which is supported by the Research Council of Norway (project number 262622), and The Royal Society through the award of funding to host the Theo Murphy Discussion Meeting “High resolution wave dynamics in the lower solar atmosphere” (grant Hooke18b/SCTM).
This research has received funding from the European Union’s Horizon 2020 Research and Innovation program under grant agreement No 82135 (SOLARNET) and No 739500 (PRE-EST). 
This research has made use of the IBIS-A archive.
We recognise the collaborative and open nature of knowledge creation and dissemination, under the control of the academic community as expressed by Camille No\^{u}s at http://www.cogitamus.fr/indexen.html.

\bibliography{sunspot_bib.bib}{}
\bibliographystyle{aasjournal}



\end{document}